\documentclass[10pt]{article}
\usepackage{graphicx}

\begin{document}

\title{Non-oscillating neutrinos in vacuum}
\author{Georgios Choudalakis}
\date{ University of Athens \\Physics Department \\ January 2004 }

\maketitle

\begin{abstract}
It is well known that matter effects affect the way neutrinos
oscillate. The amplitude of oscillation in matter can be either
enhanced, compared to the amplitude in vacuum, or suppressed,
depending on the density of matter at the vicinity of a neutrino.
Enhancement is less probable to occur than suppression.

This article demonstrates how matter effects can result into
non-oscillating neutrinos even \emph{in vacuum}.

\end{abstract}

\section{The evolution equation in matter}
We will firstly consider the case of two generations of neutrinos
in matter. In the basis of flavor, the evolution equation of a
neutrino in the 2-Dimensional flavor space takes \cite{Ach} the
form:
\begin{equation}\label{evolution}
i{\partial  \over {\partial t}}\left( {\matrix{
   {\nu _e }  \cr
   {\nu _\mu  }  \cr
 } } \right) = \left( {\matrix{
   { - A + M} & B  \cr
   B & A  \cr
 } } \right)\left( {\matrix{
   {\nu _e }  \cr
   {\nu _\mu  }  \cr
 } } \right)
\end{equation}
where $A\equiv{\frac{\Delta m^2}{4E} \cos{2\theta_0}}$,
$B\equiv{\frac{\Delta m^2}{4E} \sin{2\theta_0}}$ and
$M\equiv{\sqrt{2}G_F N_e}$. $N_e$ is the numerical density of
electrons, $G_F$ is Fermi's constant,
$\Delta{}m^2=m_{\nu_\mu}^2-m_{\nu_e}^2$ and $\theta_0$ is the
mixing angle in vacuum, the one and only parameter of the
$2\times{}2$ mixing matrix for 2 generations. At any moment, the
probability to observe the neutrino as a $\nu_e$ equals
$\left|\nu_e\right|^2$ and to observe it as a $\nu_\mu$ equals
$\left|\nu_\mu\right|^2$.

For constant $N_e$, equation (\ref{evolution}) can be solved
analytically, as we will see in the next paragraph.

\section{Constant $N_e$}

In order to solve eq.\ (\ref{evolution}) we need to diagonalize
the hamiltonian $H=\left(\matrix{  {-A+M} & {B}  \cr  {B} & {A}
\cr }\right)$. It is easy to show that, for constant $N_E$, the
oscillation probabilities are (\cite{Ach}, \cite{Bil}):
\begin{equation}\label{probs}
P_{\nu_e\rightarrow{}\nu_\mu}=P_{\nu_\mu\rightarrow\nu_e}=\sin^2{2\theta}
\sin^2{\pi \frac{L}{L_m}}
\end{equation}
where
\begin{equation}\label{tan}
\tan{2\theta}=\frac{2B}{2A-M}
\end{equation}
and
\begin{equation}\label{L_m}
L_m=\frac{2\pi}{E_A-E_B}
\end{equation}
with $E_{A,B}$ being the eigenvalues of matrix $H$, which are:
\begin{equation}
E_{A,B}=\frac{M\pm{}\sqrt{M^2+4\left(A^2+B^2-AM\right)}}{2}
\end{equation}

This solution means that neutrinos in matter of constant density
oscillate in a sinusoidal way, like in vacuum, while the amplitude
and wavelength\footnote{By saying ``neutrinos oscillation'', what
we actually mean is ``oscillatory variation of the
survival/transition probability of neutrinos as time goes by''.
Thus, amplitude and wavelength refer to the characteristics of the
sinusoidal function which expresses their survival/transition
probability.} of oscillation in matter are different from those in
vacuum. The difference results from the non-zero $M$ term in
equations (\ref{tan}) and (\ref{L_m}). When $N_e$ tends to $0$,
then $M$ also tends to $0$ and $\theta \rightarrow \theta_0$ and
$L_m \rightarrow L_0=4\pi \frac{E}{\Delta m^2}$.

An important result of this solution is that oscillation amplitude
can be either enhanced of suppressed with respect to oscillation
in vacuum, due to matter effects. This may be demonstrated by
ranging $M$ in eq.\ (\ref{tan}) from $0$ to $\infty$. For
$M\rightarrow +\infty$ we have
$\tan\left(2\theta\right)\rightarrow 0 \Rightarrow \theta
\rightarrow 90^\circ$, which means zero amplitude, suppression of
oscillation.

There is only one value of $M$ which maximizes amplitude by giving
$\theta =45^\circ$. This phenomenon is called \emph{MSW-resonance}
\cite{MSW1} and happens when:
\begin{equation}\label{MSW}
M=M_{MSW}=2A\Rightarrow \sqrt{2}G_F N_e=\frac{\Delta m^2}{2E}
\cos{2\theta_0}
\end{equation}

Regarding the wavelength of oscillation $L_m$, for zero $N_e$ it
equals $L_0$ while for $N_e \rightarrow +\infty$ it tends to zero.

\section{Variant $N_e$}

As neutrinos travel through celestial bodies and other objects,
they cross regions of extremely high and extremely low electron
densities. Thus, solving eq.\ (\ref{evolution}) for constant $N_e$
does not help much. But when $N_e$ is a function of
position\footnote{$N_e$ is a function of position, but for
neutrinos traveling at (almost) the speed of light, it is
convenient to replace $x$ with the neutrino time of flight $t$,
using Natural Units where $c=\hbar=1$.} $(x)$, then eq.\
(\ref{evolution}) gets extremely complicated and can only be
solved numerically\footnote{There are only a few exceptions, like
the well known \emph{adiabatic approximation}
(\cite{Bil},\cite{Ach}), where the evolution equation with variant
electrons density can be given an analytical approximate
solution}.

Either numerically or analytically, solutions of eq.\
(\ref{evolution}) give us the amplitudes $\nu_e(t)$ and $\nu_\mu
(t)$ as functions of time. Those amplitudes are \emph{complex}
numbers, but we usually express the solution in terms of
$\left|\nu_e (t)\right|^2\equiv{P\left(\nu_e;t\right)}$ and
$\left|\nu_\mu (t)\right|^2\equiv{P\left(\nu_\mu ;t\right)}$, as
done in eq.\ (\ref{probs}), because we have a more intuitive
understanding of those probabilities. However, as it will have
been explained by the end of this article, the complex nature of
these amplitudes is important and my lead to observable phenomena.

\begin{figure}
\centering
\includegraphics[height=100mm]{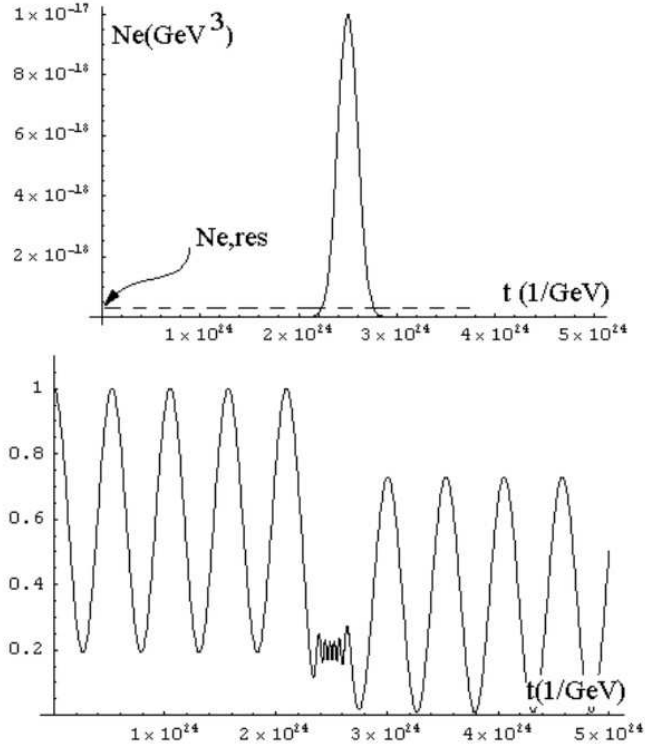}
\caption{(a): The density profile for which we solved the
evolution equation. It is a gaussian function which most of the
time exceeds $N_{MSW}$. (b): The resulting survival probability of
a neutrino which initially was a pure $\nu_e$. We can see the
region of suppression at the center of the density distribution.
We have assumed E=3 GeV, $\theta_0=32^\circ$, $\Delta
m^2=7.2\cdot10^{-5} \mbox{ } eV^2$.} \label{fig_1}
\end{figure}

When $N_e$ varies, $P\left(\nu_e ;t\right)$ and $P\left(\nu_\mu
;t\right)$ may vary in a chaotic fashion, depending on the density
profile $N_e (t)$, the initial conditions $\nu_{e,\mu} (t=0)$, the
energy $E$ of the neutrino and the input parameters $\theta_0$ and
$\Delta m^2$. An example of such an irregular oscillation can be
seen in Fig.\ \ref{fig_1}. Nevertheless, in space (time) intervals
where $N_e$ does not change very rapidly, the approximation of
constant $N_e$ holds well and oscillation patterns in those
regions look like those of eq.\ (\ref{probs}).

Another thing that should be noticed is that MSW-resonance may
occur instantly if the density profile has not been adjusted so as
to satisfy eq.\ (\ref{MSW}) for long intervals. On the other hand,
suppression is observed for $N_e \gg N_{MSW}$, which holds for a
whole range of density values, not just for a specific $N_e$. So,
if things have not been intentionally set up to be different,
suppression is a phenomenon which may last for longer and is more
probable than MSW-resonance.

\section{The complex nature of $\nu_{e,\mu}$}

Equation (\ref{evolution}) is a system of \emph{complex}
differential equations and its solutions are complex functions of
time, as stated earlier. The real and imaginary parts of $\nu_e
(t)$ and $\nu_\mu (t)$ affect the real and imaginary parts of
$\nu_e (t+dt)$ and $\nu_\mu (t+dt)$.

As for every differential equation, the solution of eq.\
(\ref{evolution}) depends on the initial condition $\nu_{e,\mu}
(t=0)$, which involves $\nu_e (0)$ and $\nu_\mu (0)$ as
\emph{complex} numbers:
\begin{equation}
\nu_\alpha = \left| \nu_\alpha \right| e^{i \phi_\alpha} \mbox{ ,
$\alpha=e, \mu$}
\end{equation}
Starting from the same $\left|\nu_\alpha\right|$ and different
$\phi_\alpha$ leads to completely different solutions
$P(\nu_\alpha,t)$, even in vacuum\footnote{As we will see soon,
there is only one case where $\phi_\alpha$ makes no difference,
and this case is the one nature always prefers.}. However, people
don't pay much attention to the arguments $\phi_\alpha$. The
reason is that neutrinos are always produced in weak vertices. All
experiments show that weak bosons $(W^\pm,Z^0)$ couple with
neutrinos in eigenstates of flavor. Thus, all neutrinos start
their journey from the initial condition:
$$
\left( \matrix{{P(\nu_e;0)} \cr {P(\nu_\mu;0)}} \right) = \left(
\matrix{{1} \cr {0}}\right) \mbox{ or } \left( \matrix{{0}\cr{1}}
\right)
$$
equivalently:
\begin{equation}\label{natural_condition}
\left( \matrix{ {\left|\nu_e\right|} \cr {\left| \nu_\mu \right|}}
\right) =\left( \matrix{{1} \cr {0}}\right) \mbox{ or } \left(
\matrix{{0}\cr{1}} \right)
\end{equation}

Of course, initial condition (\ref{natural_condition}) says
nothing about the arguments $\phi_\alpha$. They can take any
value. But if they are arbitrary and also drastically affect
$P(\nu_\alpha;t)$, then experiments should have observed extremely
messy oscillations, different for identical neutrinos with only
different $\phi_\alpha$. The reason that this is \emph{not} what
happens is that the impact of $\phi_\alpha$ on $P(\nu_\alpha;t)$
is canceled when eq.\ (\ref{natural_condition}) holds, which means
always. It can be proved analytically and also be tested
numerically.

In brief, \emph{neutrinos are eigenstates of flavor, not of mass,
and that is why they oscillate. That is also why $\phi_\alpha$
have no observable effect.}

\section{The condition of non-oscillation in vacuum}

There are at least two methods to find the condition of
non-oscillation of neutrinos in vacuum. The brute one is to solve
analytically eq.\ (\ref{evolution}) with the most general
expression of initial conditions, find the expression for
$P(\nu_\alpha;t)$ in terms of the initial conditions and then
demand that $\frac{\partial P(\nu_\alpha;t)}{\partial t}=0 \mbox{
} \forall t$. This demand imposes a condition to the initial
conditions which, if satisfied, guarantees
constancy\footnote{Constancy of $P(\nu_\alpha;t)$ does not mean
constancy of the solution $\nu_\alpha (t)$, but only of $\left|
\nu_\alpha (t) \right|$. In the complex plane, $\nu_e (t)$ and
$\nu_\mu (t)$ always rotate.} of $P(\nu_\alpha;t)$.

Constancy of $P(\nu_\alpha;t)$ is also guaranteed if the neutrino
initially is in an eigenstate of \emph{mass}, because
$E=\sqrt{p^2+m^2}$ so, assuming definite momentum $p$, definite
mass means definite energy $E$, which means constant state. So,
the second, equivalent but much more elegant method to find the
condition of non-oscillation is to demand that initially the
neutrino has definite mass.

For two generations, the mixing of flavor and mass eigenstates is
given by the expression:
$$
\left( \matrix{{\nu_e} \cr {\nu_\mu}} \right) =  \left(
\matrix{\cos{\theta_0} & \sin{\theta_0} \cr -\sin{\theta_0} &
\cos{\theta_0}}\right) \left(\matrix{\nu_1 \cr \nu_2} \right)
\Rightarrow
$$
\begin{equation} \label{mix}
\Rightarrow \left( \matrix{{\nu_1} \cr {\nu_2}} \right) =  \left(
\matrix{\cos{\theta_0} & -\sin{\theta_0} \cr \sin{\theta_0} &
\cos{\theta_0}}\right) \left(\matrix{\nu_e \cr \nu_\mu} \right)
\end{equation}
with
\begin{equation}\label{norms=1}
\left|\nu_e\right|^2+\left|\nu_\mu\right|^2=1
\end{equation}

Demanding that initially $\nu_1=0$, (\ref{norms=1}) and
(\ref{mix}) give the condition:
\begin{equation} \label{condition1}
\fbox{$\frac{\nu_e}{\nu_\mu}=\tan{\theta_0} \mbox{ and }
\left|\nu_e\right|=\frac{1}{\sqrt{1+\cot^2{\theta_0}}}$}
\end{equation}

Demanding that initially $\nu_2=0$, from (\ref{norms=1}) and
(\ref{mix}) we find:
\begin{equation} \label{condition2}
\fbox{$\frac{\nu_e}{\nu_\mu}=-\cot{\theta_0} \mbox{ and }
\left|\nu_e\right|=\frac{1}{\sqrt{1+\tan^2{\theta_0}}}$}
\end{equation}

\begin{figure}[h]
\centering
\includegraphics[height=70mm]{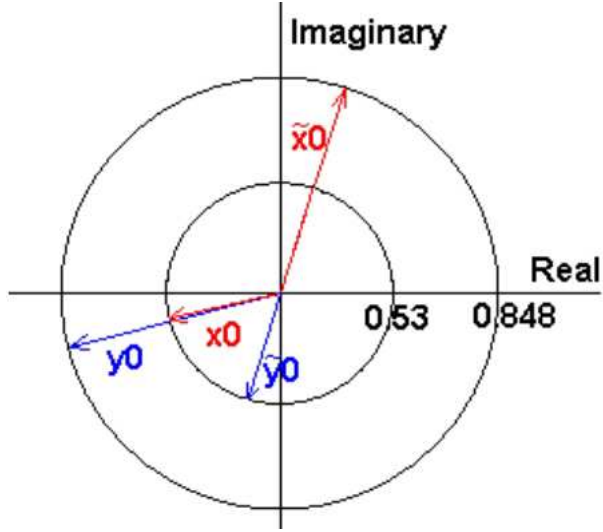}
\caption{Complex pairs like $(x_0,y_0)$ and $(\tilde x_0,\tilde
y_0)$, pointing in any direction, satisfy (\ref{condition1}) and
(\ref{condition2}) respectively, assuming $\theta_0=32^\circ$
which determines the radii of the circles. $x_0$ and $y_0$ stand
for $\nu_e$ and $\nu_\mu$.} \label{fig2}
\end{figure}

Conditions (\ref{condition1}) and (\ref{condition2}) do not only
refer to the amplitudes of $\nu_{e,\mu}$, but also to the
relationship of their arguments $\phi_\alpha$, so the complex
nature of $\nu_e$ and $\nu_\mu$ can not be marginated and consider
only $\left|\nu_\alpha\right|$. For (\ref{condition1}) to be
satisfied, $\nu_e$ and $\nu_\mu$ must be in the same direction on
the complex plane (homoparallel), while (\ref{condition2}) demands
them to point in opposite directions (antiparallel). $\theta_0$ is
the only parameter in both non-oscillation conditions. Fig.\
\ref{fig2} shows the configurations of $\nu_e$ and $\nu_\mu$ on
the complex plane, which satisfy the non-oscillation conditions.

\begin{figure}[h]
\centering
\includegraphics[height=70mm]{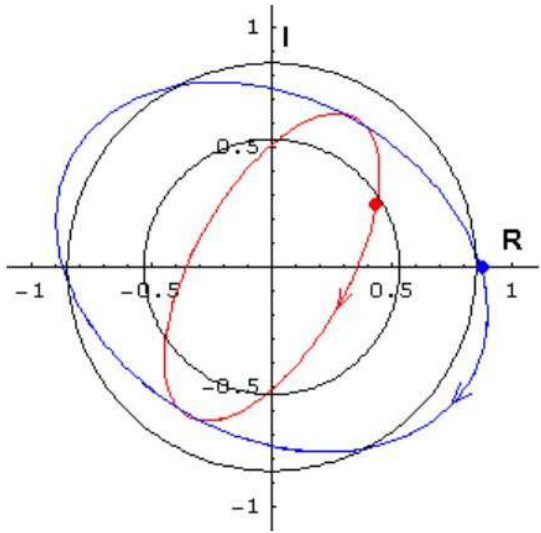}
\caption{Trajectories of x(t) and y(t) in the complex plane when
in vacuum, with initial conditions $x_0=0.5
e^{i\frac{\pi}{6}},y_0=0.866 e^{i0}$ (see the dots on the
ellipses) and $\theta_0=32^\circ$. Red ellipsis is the trajectory
of x(t) and blue is of y(t). Black circles are the same we had in
Fig.\ \ref{fig2}.}
\label{fig3}
\end{figure}

Let's examine how $\nu_e$ and $\nu_\mu$ behave as complex numbers,
when in vacuum. Let me substitute $\nu_e$ and $\nu_\mu$ with the
symbols $x$ and $y$ respectively. Analytical solving of eq.\
(\ref{evolution}) for $M=0$ gives $x$ and $y$ revolving on
ellipses on the complex plane, like in Fig.\ \ref{fig3}. Those
ellipses always have their centers at the point where axes
intersect (point 0) and are inclined with respect to the axes.
Their inclinations depend on both $x_0$ and $y_0$. The analytical
expression of those ellipses has been calculated to be:
\begin{equation} \label{revolving1}
\begin{array}{l}
Re\left[x(t)\right] =
\sqrt{\Pi_1^2+\Pi_2^2}\cos{\left(st-\arctan{\left(\frac{\Pi_2}{\Pi_1}\right)}\right)}
\cr Im\left[x(t)\right] =
\sqrt{\Pi_3^2+\Pi_4^2}\cos{\left(st+\arctan{\left(\frac{\Pi_3}{\Pi_4}\right)}\right)}
\end{array}
\end{equation}
\begin{equation} \label{revolving2}
\begin{array}{l}
Re\left[y(t)\right] =
\sqrt{B_1^2+B_2^2}\cos{\left(st-\arctan{\left(\frac{B_2}{B_1}\right)}\right)}
\cr Im\left[y(t)\right] =
\sqrt{B_3^2+B_4^2}\cos{\left(st+\arctan{\left(\frac{B_3}{B_4}\right)}\right)}
\end{array}
\end{equation}
where
\begin{equation}
\matrix{
{\Pi_1=X_{11}Re\left[C_1\right]+X_{21}Re\left[C_2\right]\mbox{, }
B_1=X_{12}Re\left[C_1\right]+X_{22}Re\left[C_2\right] } \cr
{\Pi_2=X_{21}Im\left[C_2\right]-X_{11}Im\left[C_1\right]\mbox{, }
B_2=X_{22}Im\left[C_2\right]-X_{12}Im\left[C_1\right]} \cr
{\Pi_3=X_{11}Im\left[C_1\right]+X_{21}Im\left[C_2\right]\mbox{, }
B_3=X_{12}Im\left[C_1\right]+X_{22}Im\left[C_2\right]} \cr
{\Pi_4=X_{11}Re\left[C_1\right]-X_{21}Re\left[C_2\right]\mbox{, }
B_4=X_{12}Re\left[C_1\right]-X_{22}Re\left[C_2\right]} }
\end{equation}
where
\begin{equation}
\left\{
\begin{array}{l}
{C_1
=\frac{y_0-\frac{X_{22}x_0}{X_{21}}}{X_{12}-\frac{X_{22}X_{11}}{X_{21}}}}
\cr {} \cr {C_2=\frac{x_0-X_{11}C_1}{X_{21}}}
\end{array}
\right.
\end{equation}
where
\begin{equation}
\begin{array}{l}
{\left( \matrix{X_{11}\cr
X_{12}}\right)=\left(1+\left(\frac{A-s}{B}\right)^2\right)^{-1/2}\left(\matrix{1\cr
\frac{A-s}{B}}\right)} \cr {\left(\matrix{X_{21} \cr
X_{22}}\right)
=\left(1+\left(\frac{A+s}{B}\right)^2\right)^{-1/2}\left(\matrix{1
\cr \frac{A+s}{B}}\right)}
\end{array}
\end{equation}
and
\begin{equation}
s=\frac{\Delta m^2}{4E}
\end{equation}

From (\ref{revolving1}) and (\ref{revolving2}) it is obvious that
$x(t)$ and $y(t)$ revolve with the same angular frequency $s$, so,
if they are not initially collinear in the complex plane, they
will never be. On the other hand, if we initially select $x_0$ and
$y_0$ to satisfy one of the non-oscillation conditions, then the
ellipses take the shape of the black circles of Fig. 3 and $x$ and
$y$ keep revolving coherently, being always collinear.

\section{Matter effects in complex trajectories}

Non-oscillation in vacuum would only be fiction, if matter effects
were not capable of achieving the non-oscillation condition,
because nature produces neutrinos obeying eq.\
(\ref{natural_condition}), which contradicts (\ref{condition1})
and (\ref{condition2}).

\begin{figure}
\centering
\includegraphics[width=120mm]{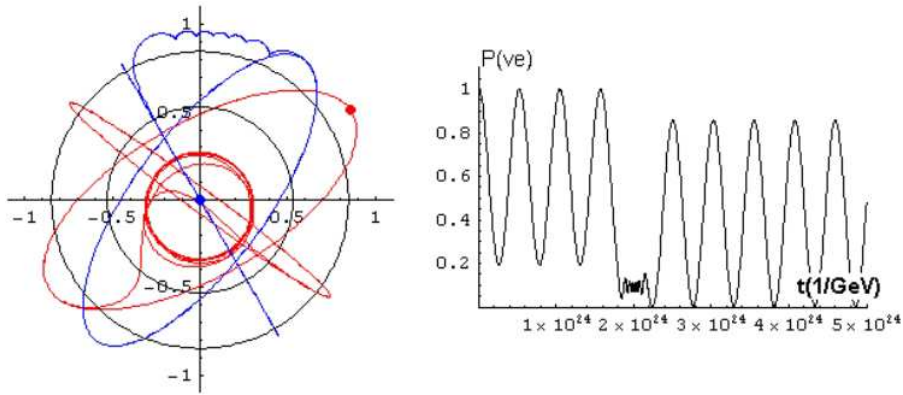}
\caption{\textbf{Right:} The survival probability of a neutrino
which initially was a $\nu_e$ with $x_0=1\cdot
e^{i\frac{\pi}{6}},y_0=0$ and crosses a region of high matter
density, such as a star. We have assumed a gaussian $N_e(t)$ of
the form of eq.\ (\ref{Ne_t}) with $\rho_0=10^{-17} \mbox{ }
GeV^3$, $t_0=2\cdot 10^{24} \mbox{ } GeV^{-1}$ and $\sigma=10^{23}
\mbox{ } GeV^{-1}$. Suppression is obvious around $t=t_0$, as
$N_e$ highly exceeds $N_{MSW}$. The oscillation is drastically
modified by the passing through the star. \textbf{Left:} The
complex trajectories of x(t) (red) and y(t) (blue). Before
entering the star, x(t) circulates along the red ellipsis (red and
blue dots indicate $x_0$ and $y_0$ respectively) and y(t) along
the blue ellipsis which is degenerated into a line for those
(naturally default) initial conditions.} \label{fig4}
\end{figure}

Numerical solution of eq.\ (\ref{evolution}) with variant $N_e$
has revealed that the trajectories of $x$ and $y$ in the complex
plane get extremely complicated (see Fig. \ref{fig4}) and
\emph{can} happen to satisfy one of the non-oscillation
conditions, even instantaneously. In other words, a neutrino in
matter may transiently get into a mass eigenstate.

From the left part of Fig. \ref{fig4} we realize that, as density
increases and then decreases, $x(t)$ and $y(t)$ drift away from
the ellipsoidal trajectories they initially followed in vacuum.
When density becomes zero again, they return to elliptical
trajectories which are different from those they initially
followed when in vacuum. The reason is that the neutrino returned
to vacuum from different initial conditions, i.e.\ the $x$ and $y$
it had at the moment it exited the matter region.

The most critical interval is while the neutrino lies in the
middle of the matter region, where its oscillation is highly
suppressed. Then, $y(t)$ moves along the curly blue line, around
$y(t) \simeq something + i\cdot 0.9$. At the same time, $x(t)$
rapidly spins along the red circles of radius $\sim0.3$. In what
trajectories they will end up depends on \emph{when} the spinning
of $x(t)$ and the curling of $y(t)$ will stop, i.e.\ when $N_e$
will decline.

If $N_e$ rises appropriately and diminishes at the right time,
then it is possible to make $x(t)$ and $y(t)$ satisfy (or almost
satisfy) one of the non-oscillation conditions. After methodically
trying several initial conditions (always obeying eq.\
(\ref{natural_condition})) and several $N_e$ shapes, I found one
example of such a `flavor lens'\footnote{Actually, trial-and-error
method would have very hardly lead to any findings. I had to
implement a `trick': At each time step of the numerical solution I
would check if any non-oscillation condition were almost
satisfied. If it were, then I would `cut' the density distribution
and set it equal to zero from there on. In this way, the neutrino
would be dropped back to vacuum on the right time.}. The neutrino
was assumed to initially have $x_0=1\cdot e^{i0}$, $y_0=0$ and
energy E=3 GeV. It has also been assumed that $\theta_0=32^\circ$
and $\Delta m^2=7.2\times 10^{-5} \mbox{ } eV^2$.

The density profile found can be seen in Fig. 5, together with the
oscillation of survival probability of a $\nu_e$ passing through
it. $N_e(t)$ rises according to the function:
\begin{equation}\label{Ne_t}
N_e(t)=\rho_0 \exp{\left(-\frac{(t-t_0)^2}{2\sigma^2}\right)}
\end{equation}
with
$$
\left\{
 \begin{array}{l}
 {t_0=4\cdot 10^{24} \mbox{ } GeV^{-1}}\cr
 {\rho_0=10^{-16} \mbox{ } GeV^3}\cr
 {\sigma=10^{24} \mbox{ } GeV^{-1} }\cr
 {N_e(t) = 0 \mbox{ for } t>3.03\cdot 10^{24} \mbox{ } GeV^{-1} }
 \end{array}
\right.
$$

The fact that I found this specific density profile does not mean
that it is the only `flavor lens' for this kind on neutrinos. The
same distribution, with its center $t_0$ translated by an integer
number of vacuum wavelengths $L_0$, would have the same result.
Whether a density profile is a flavor lens or not depends on its
position with respect to neutrino's production point, on
neutrino's energy, on its shape and on the initial conditions of
the neutrino $(x_0,y_0)$.

It has been tested and confirmed that $\phi_\alpha$ do not affect
the oscillation through the flavor lens, as long as
(\ref{natural_condition}) was initially true. In the same way that
(\ref{natural_condition}) `hides' the impact of $\phi_\alpha$ in
oscillations in vacuum, it also also prevents the role of
$\phi_\alpha$ from being revealed when matter is present. So, the
same oscillation would occur whether $(x_0=1\cdot e^{i0},y_0=0)$
or $(x_0=1\cdot e^{i\pi/3},y_0=0)$ etc.\ (see Fig. 8).
\begin{figure}
\centering
\includegraphics[height=50mm]{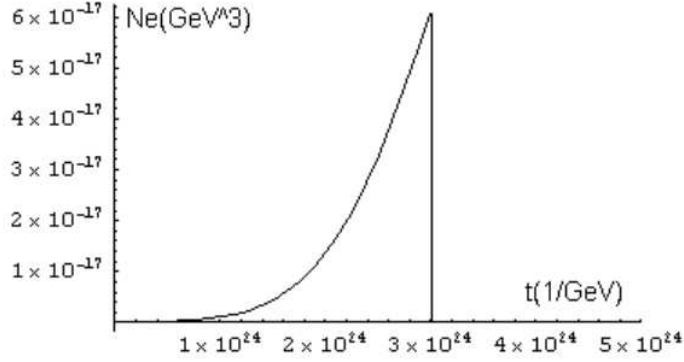}
\caption{A density profile which acts as a flavor lens for $\nu_e$
of E=3 GeV, provided that $\theta_0=32^\circ$ and $\Delta
m^2=7.2\cdot 10^{-5}$ $eV^2$.} \label{fig5}
\end{figure}
\begin{figure}
\centering
\includegraphics[height=70mm]{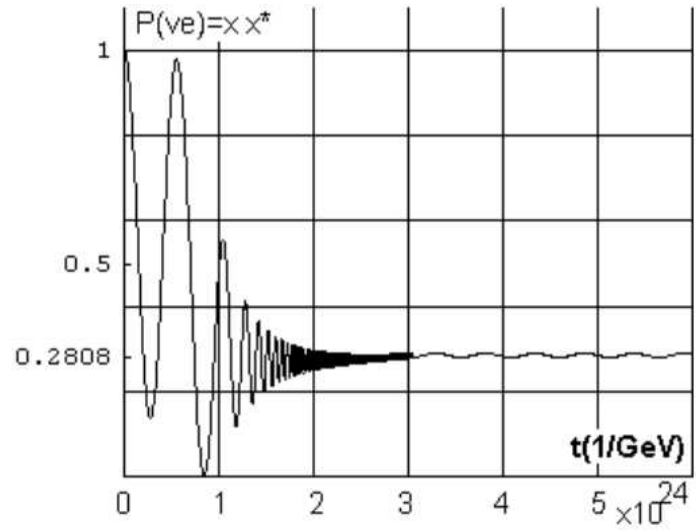}
\caption{Survival probability for a $\nu_e$ of E=3 GeV, as it
passes through the flavor lens of Fig.\ \ref{fig5}. When it has
emerged from the matter region it is almost non-oscillating. The
fact that it undergoes very slight oscillation is because the
non-oscillation condition was not exactly satisfied when matter
dropped to zero, but was almost satisfied.} \label{fig6}
\end{figure}
\begin{figure}
\centering
\includegraphics[height=70mm]{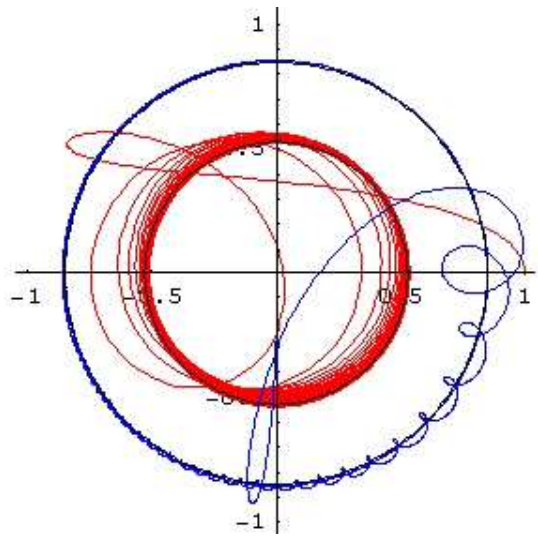}
\caption{The trajectories of x(t) (red) and y(t) (blue) as a 3 GeV
neutrino with $x_0=1\cdot e^{i0},y_0=0$ passes through the flavor
lens of fig.\ \ref{fig5}.} \label{fig7}
\end{figure}
\begin{figure}
\centering
\includegraphics[height=70mm]{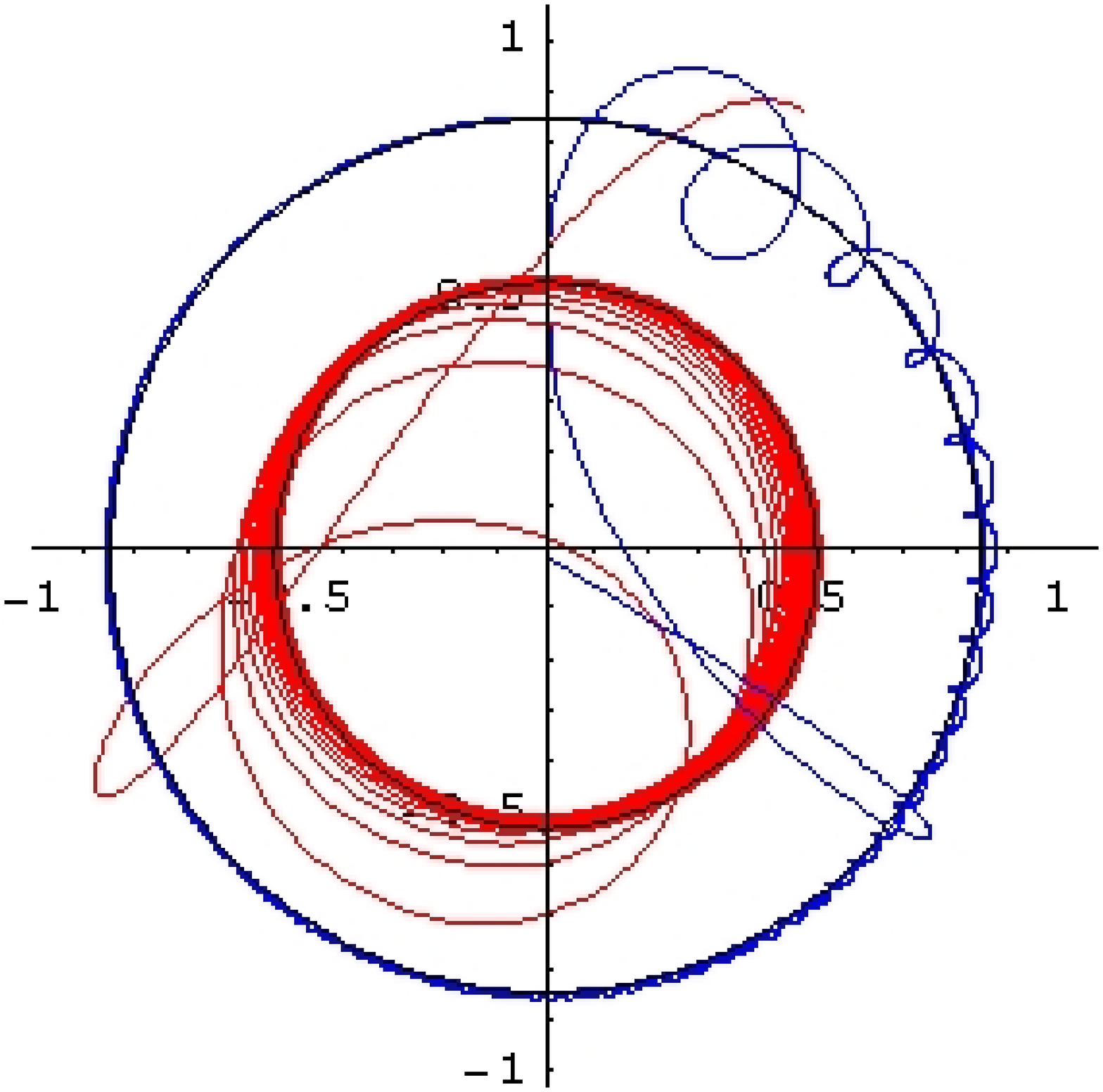}
\caption{The same like in fig. \ref{fig7}, but with initial
conditions $x_0=1\cdot e^{i\frac{\pi}{3}},y_0=0$. Both
trajectories are rotated by $\pi/3$, but their behavior is the
same as fig. \ref{fig7}, so the survival probability remains the
one in fig. \ref{fig6}.} \label{fig8}
\end{figure}

\section{3 Generations}

In 3 generations things do not differ much in principle. The
equation of evolution is a system of 3 differential equations,
where there is a matter term equivalent to M we had in 2
generations \cite{Ach}.

Demanding that a neutrino is at an eigenstate of mass $\nu_i$,
instead of an eigenstate of flavor $\nu_\alpha$, we derive the
non-oscillation condition in 3 generations.

Starting with the notation:
\begin{equation}
\nu_\alpha=\sum\limits_{i=1}^{3} U_{\alpha i}\nu_i \mbox{ }
\Rightarrow \nu_i=\sum\limits_{\alpha=e,\mu,\tau} {U_{\alpha
i}^\ast\nu_\alpha}
\end{equation}
and respecting the completion condition:
\begin{equation}
\left|\nu_e\right|^2+\left|\nu_\mu\right|^2+\left|\nu_\tau\right|^2=1
\end{equation}
we find that to have an eigenstate of mass, i.e.\ to have
$\nu_{i''}=1, \nu_{i'}=\nu_i=0$, the condition is:
\begin{equation}
\left\{
\begin{array}{l}
{\left|\nu_\mu\right|=1/\sqrt{1+\left|A\right|^2+\left|B\right|^2}}\cr
{\nu_e=A\nu_\mu}\cr {\nu_\tau=B\nu_\mu}
\end{array}
\right.
\end{equation}
where
\begin{equation}
A=\frac{\frac{U_{\mu i}^\ast U_{\tau i'}^\ast}{U_{\tau
i}^\ast}-U_{\mu i'}^\ast}{U_{e i'}^\ast-\frac{U_{ei}^\ast U_{\tau
i'}^\ast}{U_{\tau i}^\ast}} \mbox{ , }
B=\left(-\frac{U_{ei}^\ast}{U_{\tau i}^\ast}A-\frac{U_{\mu
i}^\ast}{U_{\tau i}^\ast} \right)
\end{equation}

In 3 generations, the non-oscillation condition is triple, one for
each eigenstate of mass, and depends only on the tree mixing
angles of the PMNS matrix U.

Non oscillation condition in 3 generations is much more demanding
than in 2 generations. It demands three complex numbers to be
collinear in the complex plane, to point in the right directions
and to be of the right sizes. However, achieving its satisfaction
with an appropriate density profile must not be impossible, though
it is not easy to find an example.

\section*{Conclusions}

Matter effects in neutrinos oscillation have been discussed. The
complex character of $\nu_\mu$ and $\nu_e$ has been emphasized, as
density $N_e$ affects their trajectories in the complex plane,
potentially leading neutrinos, transiently, to mass eigenstates.

It has been demonstrated how an appropriate $N_e(t)$ can transform
a regular $\nu_e$ into one which does not oscillate in vacuum.

Finally, the non-oscillation condition in 3 generations has been
presented.

\section*{Acknowledgements}
I would like to thank my supervisor at the University of Athens,
Professor George Tzanakos for the discussions we had about this
subject, and my colleague Panos Stamoulis for the same reason. I
would also like to thank Professor Carlo Giunti for reading this
draft, suggesting a few corrections and expressing his opinion to
me.

\end{document}